\documentclass[twocolumn]{aastex7}
\usepackage{amsmath}
\usepackage{booktabs}
\usepackage{fontawesome}
\usepackage{bm}
\usepackage{amsbsy}
\usepackage{fontawesome}
\usepackage{threeparttable}

\usepackage{soul, CJK}
\usepackage[utf8]{inputenc}

\usepackage{xspace}
\DeclareFixedFont{\ttb}{T1}{txtt}{bx}{n}{9} 
\DeclareFixedFont{\ttm}{T1}{txtt}{m}{n}{9}  
\usepackage{color}
\definecolor{deepblue}{rgb}{0,0,0.5}
\definecolor{deepred}{rgb}{0.6,0,0}
\definecolor{deepgreen}{rgb}{0,0.5,0}
\usepackage{listings}
\usepackage{gensymb}

\newcommand\pythonstyle{\lstset{
language=Python,
basicstyle=\ttm,
otherkeywords={self},             
keywordstyle=\ttb\color{deepblue},
emph={MyClass,__init__},          
emphstyle=\ttb\color{deepred},    
stringstyle=\color{deepgreen},
frame=tb,                         
showstringspaces=false            %
}}
\lstnewenvironment{pyt}[1][]
{
\pythonstyle
\lstset{#1}
}
{}

\newcommand{\R}{\mathbb{R}}
\newcommand{\N}{\mathbb{N}}

\newcommand{\g}{\,|\,}

\usepackage{listings}

\graphicspath{{./}{figures/}}

\shorttitle{A Mass-Ratio Desert Around Low-Mass Stars}
\shortauthors{Zhang}

\begin{document}
\begin{CJK*}{UTF8}{gkai}

\title{The Brown-dwarf Desert Persists as a Mass-ratio Desert around Low-mass Stars}

\correspondingauthor{Keming Zhang}

\author[0000-0002-9870-5695]{Keming Zhang (张可名)}
\altaffiliation{NASA Hubble Fellowship Program (NHFP) Sagan Fellow}
\affiliation{Kavli Institute for Astrophysics and Space Research, Massachusetts Institute of Technology, MA 02139, USA}
\affiliation{Department of Astronomy and Astrophysics, University of California, San Diego, CA 92093, USA}
\email[show]{kemingz@mit.edu}

\begin{abstract}
Sun-like stars are known to host a paucity of brown dwarf companions at close separations.
Direct imaging surveys of intermediate-mass stars have suggested that the brown dwarf desert may be fundamentally a feature in the mass ratio.
Microlensing surveys provide a unique opportunity to investigate the nature of this desert around low mass stars, as microlensing hosts have typical masses of 0.05--0.8 $M_\odot$.
Here, we perform a statistical analysis of homogeneously selected binary-lens microlensing events in the literature, and identify a companion mass-ratio desert at $0.02\lesssim q\lesssim0.05$ and projected separations around 1--5 au.
We derive a statistically significant truncation to the giant-planet mass-ratio distribution at $q\simeq0.02$, above which the occurrence rate density drops by approximately an order of magnitude.
Due to the possibility that the small number of detected companions in this desert orbit white dwarfs, this mass-ratio desert may be closer to being completely dry around main-sequence stars.
Below this desert, we derive a giant-planet ($q>4\times10^{-4}$) occurrence rate density of $5.8\pm0.8$\% per decade of projected separation.
Our analysis furthers the hypothesis that the brown dwarf desert is fundamentally a feature in the mass ratio, separating distinct populations of planetary and non-planetary companions that are likely formed via core accretion and gravitational instability, respectively.
The persistence of this desert across stellar types suggests that both gas-giant planets and sub-stellar companions form in a scale-invariant fashion, with planets growing to a maximum of approximately 2\% their host masses.
\end{abstract}


\section{Introduction}
Since the late 1990s, radial velocity surveys have uncovered a dearth of brown-dwarf companions (13--80 $M_{\rm Jup}$) within $\sim$5 au of Sun-like stars (e.g., \citealt{marcy_planetary_1998,grether_how_2006}), commonly referred to as the brown dwarf desert.
By investigating how this desert may persist for other types of stellar hosts, one may empirically infer the fundamental principles underlying the formation of stellar and planetary companions.
A recent high-contrast imaging survey \citep{duchene_low-mass_2022} of 169 intermediate-mass ($1.75$-$4.5\,M_\odot$) primaries achieved sub-stellar sensitivity for three-quarters of their targets, but found no companions below $0.3 M_\odot$ ($q\lesssim0.1$). When their sample is combined with known BD companions to A--F stars in the literature, the resulting distribution revealed a low-mass stellar companion desert spanning $0.02\lesssim q\lesssim0.05$, matching the brown dwarf desert around solar-type stars in the mass ratio, albeit at wider separations of 20--1000 au.

Microlensing surveys are ideally suited to examine an analogous desert around low-mass stars, as they directly measure the mass ratio, are agnostic to companion brightness, and have typical host masses between 0.05--0.8 $M_\odot$, approximately half of which are M dwarfs. The peak microlensing sensitivity to giant planets and brown dwarf companions lies at projected separations of 1--5 au and is centered on the host Einstein radius (typically 2--3 au).
Statistical analyses of an early microlensing sample have suggested a minimum companion frequency at $q\simeq0.01$, with either side described by distinct power laws \citep{shvartzvald_frequency_2016}.
However, the potential existence of an extended mass-ratio desert with well-defined edges has not yet been evaluated.
A primary barrier to this investigation is that existing homogeneously selected microlensing samples are focused on planetary companions \citep{suzuki_exoplanet_2016,zang_microlensing_2025}, which commonly adopt a cutoff of $q<0.03$ \citep{bond_ogle_2004}.
As a result, the upper edge of a possible desert ($q\simeq0.05$) has evaded the attention of current analysis, whereas a possible lower edge near $q\simeq0.02$ is often smeared out due to the choice of binning.

\section{Sample Selection}

We begin by examining the two largest homogeneously selected samples of microlensing companions: the Microlensing Observations in Astrophysics (MOA-II) 2007--2012 sample \citep{suzuki_exoplanet_2016}, and the Korea Microlensing Telescope Network (KMTNet) 2016--2019 sample \citep{zang_microlensing_2025}. Both adopted a formal selection threshold at $q<0.03$.
As such, we first attempt to recover $q>0.03$ companions associated with the two samples.

The MOA-II survey searched for planetary companions by fitting a binary-lens model to 1474 microlensing events satisfying their selection criteria \citep{suzuki_exoplanet_2016}, with a total of 22 planets satisfying $q<0.03$ identified.
Among them, 18 planets had been previously identified by eye, suggesting an approximately 80\% by-eye search completeness\footnote{In this work, the term ``completeness'' exclusively refers the search completeness, which is the fraction of detectable companions that are eventually identified and published.}.
Their sample included one planet within the putative mass-ratio desert: MOA-2011-BLG-322 (\citealt{shvartzvald_moa-2011-blg-322lb_2014}; $q=0.028$).
Beyond this homogeneous analysis, two other MOA-II events with $q>0.03$ were independently identified by eye and published: MOA-2007-BLG-197 ($q = 0.047$; \citealt{ranc_moa-2007-blg-197_2015}) and MOA-2010-BLG-073 ($q=0.065$; \citealt{street_moa-2010-blg-073l_2013}), which are near or above the putative desert upper edge.
The identification of these events likely reflects a continuation of the by-eye-search completeness.

In comparison, the KMTNet survey only modeled a subset of their events flagged by their match-filtering algorithm called AnomalyFinder \citep{zang_systematic_2021} as possibly planetary.
To ensure completeness for their $q<0.03$ planetary sample, the flagged events are first modeled using a binary-lens model on their standard pipeline photometry, and those with $q<0.06$ solutions are then sent for re-reduction for improved photometry and subsequently remodeled. 
A total of 112 planets with final unambiguous mass ratios of $q<0.03$ have been identified over the 2016--2019 seasons \citep{gui_systematic_2024}.
Two events were detected in the putative mass-ratio desert: OGLE-2016-BLG-1635 ($q=0.025$) and OGLE-2016-BLG-0263 ($q=0.03$).

\begin{figure}[t]
    \centering
    \includegraphics[width=\linewidth]{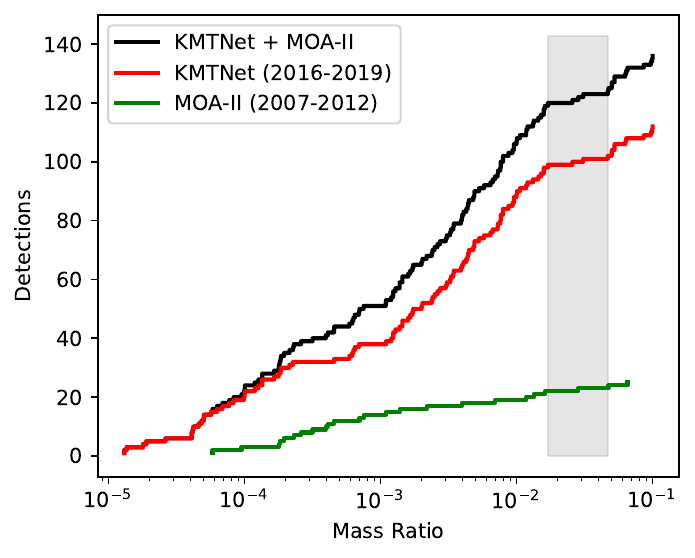}
    \caption{Cumulative number of detected microlensing companions with mass ratios smaller than each value on the horizontal axis, shown on a logarithmic scale. Events with severe degeneracies in the mass ratio are excluded from this plot. The gray shade indicates a dearth of detections between mass ratios of $0.017<q<0.047$.}
\end{figure}

Although events with $q>0.03$ are excluded from their statistical sample, the KMTNet group generally adopted a policy to report on all events that satisfied the initial $q<0.06$ criteria on the pipeline data, regardless of the final mass ratio based on the re-reduced data \citep{gould_systematic_2022}.
Among the 2016--2019 seasons, there exist a total of 5 events that passed the initial $q<0.06$ threshold but ended up with at least one solution within $0.03<q<0.1$ based on the re-reduced data \citep{zang_systematic_2022, shin_systematic_2023}.
Notably, all 5 events have final mass ratios of $q>0.047$, with three events clustered at $0.047<q<0.053$, and one with degenerate solutions of $q=0.047$ and $q=0.057$ (see Appendix A).
As noted by \cite{zang_systematic_2022,jung_systematic_2023}, the inferred mass ratios typically differed by less than 50\% based on the online and re-reduced data.
Therefore, the search completeness for this extended sample should remain substantial out to these additional detections at $q\simeq0.05$.
Beyond the AnomalyFinder sample, \cite{han_brown_2022,han_probable_2023} independently identified six brown dwarf candidates with $q<0.1$ within the 2016--2019 KMTNet seasons, with the lowest-mass ratio event being OGLE-2017-BLG-0614 ($q=0.05$).

\section{A companion mass-ratio desert}
As illustrated in Figure 1, the KMTNet and MOA-II samples with the additional $q>0.03$ events suggest a dearth of detected companions with mass ratios of $0.017<q<0.047$.
Between the two samples, there is a pile-up of five events within a narrow range of $0.047<q<0.053$, which contrasts sharply with the absence of detections over $0.03<q<0.047$, an interval 4 times wider in log-space.

While both samples are nominally complete for $q<0.03$, some assumptions must be made regarding their search completeness beyond this limit.
As far as the extended KMTNet sample is concerned, the search completeness is expected to decline monotonically over $0.03<q<0.05$, due to the increasing fraction of companions expected to fail its initial $q<0.06$ cutoff on the pipeline data.
On the other hand, the planet detectability (i.e., survey sensitivity; see below) changes by a modest $\sim$20\% between $0.03<q<0.05$.
Therefore, it is difficult to attribute the observed pile-up to selection effects.
These arguments support the identification of a desert upper edge at $q\simeq0.05$. In Section 4, we suggest hints of another frequency increase around $q=$ 0.06--0.07.

By comparison, it is possible to examine the desert lower edge in a statistically robust manner, as it is expected to lie within the formal selection criteria of $q<0.03$ for both the KMTNet and MOA-II samples.
As a first step, we derive the occurrence rate density for the combined KMTNet and MOA-II samples in equal log-spaced bins over $-5<\log(q)<-1.75$. Coincidentally, the final bin edge $q=10^{-1.75}\simeq0.0177$ is only slightly above the apparent desert lower edge, so we adopt this default choice of binning, and include a final bin of $0.0177<q<0.047$ reflecting the mass-ratio desert.
We conservatively assume that the KMTNet sample completeness uniformly drops by 50\% between $q=0.03$ and $q=0.047$. We adopt the MOA-II sample completeness as 80\% over the same range, based on the aforementioned by-eye search completeness.

Due to the very large Poisson uncertainty arising from only three detected companions in the final desert bin, the assumed search completeness for $q>0.03$ does not constitute a dominant source of uncertainty.
To illustrate this point, note that the pile-up of detected events at $q\simeq0.05$ indicates that the completeness cannot be zero here.
Instead, if we assume that the KMTNet completeness gradually dropped to zero over $0.03<q<0.047$, then the average completeness would be 50\% over this range, which compares to a 75\% average completeness under the current assumptions.
Therefore, the above hypothetical scenario of zero completeness at $q=0.047$ would only reduce the effective survey sensitivity by 15\% over $0.018<q<0.047$, which is negligible given the large Poisson uncertainty from the mere three detections in this bin.

\begin{figure}[t]
    \centering
    \includegraphics[width=\linewidth]{power-law.pdf}
    \caption{Mass-ratio distribution of microlensing companions from the KMTNet (2016--2019) and MOA-II (2007-2012) survey samples, extended out to $q=0.047$. The black points (left axis) show occurrence rate densities in bins of the mass ratio, shown as the posterior median and 16--84 percentile interval under a Jeffreys prior. The solid orange line represents the best power-law fit over $-3.4<\log(q)<-1.75$, with the shaded region indicating the $1\sigma$ confidence interval. The light shade shows extrapolation to the mass-ratio desert. The histogram (right axis) shows the number of detections in each bin.}
\end{figure}

\begin{table*}[t]
  \centering
  \caption{Microlensing events in the extended MOA-II and KMTNet samples that are unambiguously in or near the mass-ratio desert.
  Here, $q$ is the companion-to-host mass ratio and $s$ is the projected separation in units of the lens angular Einstein radius (typically 2--3 au).
  Events detected between 2007--2012 belongs to the MOA-II sample, and those detected between 2016--2019 belong to the KMTNet sample, regardless of the event names. When multiple solutions exist, the first row lists the best-fit light-curve model, with $\Delta\chi^2$ indicating the relative goodness-of-fit for the alternative models. Two degenerate models for OGLE-2016-BLG-0263 are merged due to their similarity. No uncertainties or projected separations were reported for three events.}
  \renewcommand{\arraystretch}{1.1}
  \begin{tabular}{lcc l}
    \hline
    Event                       & $q$ ($10^{-2}$)   & $s$            & Reference \\
    \hline
    MOA-2012-BLG-006            & $1.65\pm0.06$     & $4.41\pm0.07$  & \cite{poleski_companion_2017} \\
    \hline
    OGLE-2019-BLG-0954          & $1.70\pm0.20$     & $0.74\pm0.01$  & \cite{han_three_2021} \\
    \hline
    OGLE-2016-BLG-1635 & $2.55\pm0.34$     & $0.59\pm0.01$  & \cite{shin_systematic_2023} \\
    ($\Delta\chi^2=7.7$)          & $2.48\pm0.31$     & $1.68\pm0.04$  &  \\
    \hline
    MOA-2011-BLG-322   & $2.80\pm0.10$     & $1.82\pm0.01$  & \cite{shvartzvald_moa-2011-blg-322lb_2014} \\
    \hline
    OGLE-2016-BLG-0263 & $3.06\pm0.08$     & $4.72\pm0.12$  & \cite{han_ogle-2016-blg-0263lb_2017} \\
    ($\Delta\chi^2=7.8$) & $2.97\pm0.09$     & $4.86\pm0.15$  & \cite{han_ogle-2016-blg-0263lb_2017} \\
    \hline
    MOA-2007-BLG-197            & $4.73\pm0.20$     & $1.13\pm0.00$  & \cite{ranc_moa-2007-blg-197_2015} \\
    \hline
    OGLE-2019-BLG-1067          & $4.7$             & --             & \cite{zang_systematic_2023} \\
    \hline
    KMT-2019-BLG-0814 &5.0&--&\cite{zang_systematic_2022}\\
    \hline
    OGLE-2017-BLG-0614&$5.0\pm0.6$&$1.84 \pm0.07$&\cite{han_brown_2022}\\
    ($\Delta\chi^2=1.7$)&$4.9\pm0.6$&$0.53\pm0.02$\\
    \hline
    KMT-2019-BLG-3301&5.3&--&\cite{zang_systematic_2022}\\
    \hline
  \end{tabular}
\end{table*}

We derived the occurrence rate density for the combined KMTNet and MOA-II samples by dividing the number of unambiguous planet detections in each bin by the average survey sensitivity within that bin and for the two surveys combined.
The survey sensitivity as a function of $q$ is defined as the expected number of companion detections per decade of $q$ under a fiducial occurrence rate of one planet per decade of $s$ and $q$ (cf. Equation 3), where $s$ is the projected separation in units of the lens Einstein radius, given the planet detection algorithm and the underlying set of microlensing events. Therefore, the KMTNet and MOA-II samples can be combined into a homogeneous sample by adding their survey sensitivities.
The KMTNet survey sensitivity is published in steps of 0.1 dex in the mass ratio up to $q\simeq0.04$ \citep{zang_microlensing_2025}, which we linearly extrapolate to $q=0.047$.
The MOA-II survey sensitivity is published in steps of 0.5 dex in the mass ratio up to $q\simeq0.063$ \citep{suzuki_exoplanet_2016}.
The survey sensitivities are scaled down for $q>0.03$ according to the assumed completeness for KMTNet and MOA-II, and interpolated onto a common grid.

As shown in Figure 2, the resulting sensitivity-corrected companion frequencies exhibit a pronounced deficit in the final desert bin.
Based on the three companions detected in this desert between the KMTNet and MOA-II samples, we infer the desert occurrence rate density to be $1.8^{+1.2}_{-0.8}\times10^{-3}$ per decade of $s$ and $q$, which is lower by a factor of 8 (with large Poisson uncertainty) than the adjacent bin ($q\sim0.01$), and by a factor of 28.5 when compared to the more abundant Jovian analogs ($q\sim0.001$).
However, it is important to note that the current sample selection has excluded events with severe degeneracies in the mass ratio.
The excluded events account for $<15\%$ of the total planet detections in the KMTNet sample \citep{zang_microlensing_2025}. Nevertheless, given the very small number of detected companions in the desert bin, the excluded events could potentially lead to a large underestimation of the true desert abundance.

We identify three events that failed the KMTNet selection criterion due to ambiguities in the mass ratio, but with at least one plausible solution within the putative desert (Table A1). No such events are found within the MOA-II sample. These ambiguous events have degenerate models corresponding to ice-giant planets ($q<0.0003$), gas-giant planets ($q>0.0003$), or stellar binaries ($q>0.1$).
An additional event KMT-2017-BLG-2197 that passed the KMTNet criterion has two mass-ratio solutions straddling $q\simeq0.02$, each with relatively large uncertainties.
An equal weighting of the two solutions leads to a 95\% confidence interval of $q=$ 0.014--0.030, so we regard it as an ambiguous event as well.
Note that even if all four events were counted as desert events, the desert occurrence rate density is still lower by a factor of 12.2 compared to the Jovian analogs ($q\sim0.001$).

By comparison, the occurrence rate density of both ice-giant planets (Figure 2) and stellar binaries \citep{ward-duong_m-dwarfs_2015,shvartzvald_frequency_2016} is higher than that of gas-giant planets by approximately an order of magnitude.
Therefore, ice giant planet and stellar binary solutions are favored \textit{a priori} over the desert solution by approximately two orders of magnitude, with gas giant solutions favored by one order of magnitude.
After considering the observed evidence specific to each ambiguous event, we find that none of them are likely to reside in the mass-ratio desert (see Appendix B).
The nature of KMT-2017-BLG-2197 remains ambiguous, so we exclude it from our statistical analysis.

\section{A truncated mass-ratio distribution}

As illustrated in Figure 2, the planet mass-ratio distribution from the combined KMTNet and MOA-II samples exhibits a substantial drop at $q\gtrsim3\times10^{-4}$ ($\log(q)\gtrsim-3.5$), which corresponds to Neptune mass planets around $\lesssim0.17M_\odot$ hosts, or Saturn mass planets around $\lesssim0.95M_\odot$ hosts.
Therefore, this apparent ``cliff'' in the mass-ratio distribution is consistent with a sharp drop in the planet frequency above approximately Neptune mass.
Note that the same feature is not informative of a possible dearth of planets below Saturn mass, based on the scenario of runaway gas accretion \citep{ida_toward_2004}.

Below this apparent cliff, the planet frequency appears to peak at $\log(q)\simeq-3$, but the decrease for $-3.5<\log(q)<-3$ is not statistically significant. By comparison, the KMTNet sample alone has suggested a larger drop in the planet occurrence immediately below $q=0.001$, which is referred to as the sub-Saturn desert \citep{zang_microlensing_2025}, whereas the MOA-II sample has suggested an increase instead \citep{suzuki_exoplanet_2016}. This difference may be attributed to a combination of small-number statistics and systematics specific to each analysis.
By relying on the combined KMTNet and MOA-II samples, we mitigate potential systematics specific to each survey, while noting that the divergence regarding the sub-Saturn desert does not significantly impact the analysis of the companion mass-ratio desert.

Following standard practices (e.g., \citealt{suzuki_exoplanet_2016,zang_microlensing_2025}), we fit a power-law mass-ratio function to the combined KMTNet and MOA-II samples over $-3.4<\log(q)<-1.75$ (solid orange shade in Figure 2)
\begin{equation}
    f(q)=\dfrac{d^2N_{\rm planet}}{d\log(s)\,d\log(q)}=A\cdot \left(\dfrac{q}{q_0}\right)^\gamma,
\end{equation}
where the pivot point $q_0=0.003$ is chosen close to the geometric mean within the fitted range, which allows for the parameters $A$ and $\gamma$ to be largely uncorrelated. The lower limit of $q=10^{-3.4}\simeq4\times10^{-4}$ is chosen as a threshold for gas-giant planets, which restricts contamination by Neptune-mass planets to the lowest mass hosts with $<0.1M_\odot$, while preserving Saturn-mass planets around $<0.7M_\odot$ hosts.

We fit for the power-law mass-ratio function by maximizing the unbinned Poisson likelihood
\begin{equation}
    \mathcal{L} = e^{-N_{\mathrm{exp}}} \prod_{i=1}^{N} f(q_i)\,S(\log q_i),
\end{equation}
where $N$ is the total number of planets in the sample, $S(\log q_i)$ is the survey sensitivity at each observed mass ratio, and $N_{\mathrm{exp}}$ is the expected number of planets detected, given by
\begin{equation}
N_{\mathrm{exp}} = \int_{\log q_{\rm min}}^{\log q_{\rm max}}f(q)\,S(\log q)\,d\log q.
\end{equation}

We assume uniform priors on ${A, \gamma}$ and sample their posterior distribution using the \texttt{emcee} package \citep{foreman-mackey_emcee_2013}, resulting in 
\begin{align}
    A&=(3.17\pm0.36)\times10^{-2}\\
    \gamma&=-0.32\pm0.11.
\end{align}

By comparing the expected number of detections under the best-fit power law, and the actual number of detections for the seven discrete bins over $-3.4<\log(q)<-1.75$, we find a reduced chi-squared $\chi^2_{\nu}=0.86$ under Poisson statistics, which indicates that the power law parameterization is sufficient.
By integrating the inferred power law over the same mass-ratio range, we derive a giant-planet occurrence rate density of $5.8\pm0.8$\% per decade of projected separation at the peak microlensing sensitivity of 2--3 au, which is largely consistent with M-dwarf giant planet occurrence rates derived from radial velocity surveys (e.g., \citealt{mignon_radial_2025}).

By extrapolating this power law to the mass-ratio desert, we should expect $27^{+7}_{-6}$ detections between $0.018<q<0.047$, in contrast to the three definitive detections.
The null hypothesis that the mass-ratio desert follows the same power law is rejected at the 4.1-$\sigma$ level (one-sided; $p=2\times10^{-5}$).
As a robustness check, if there were indeed a drop in the planet frequency below Saturn mass, it would not substantially affect the mass-ratio distribution for $q>0.002$, assuming most microlensing hosts are more massive than $0.14M_\odot$. By restricting the power-law fit to planets satisfying $q>0.002$, we derive a power-law index of $\gamma=-0.39\pm0.22$, which is steeper than but consistent with the previously inferred index.
An extrapolation to the mass-ratio desert predicts $24^{+10}_{-7}$ detections, with the null hypothesis rejected at the 3.3-$\sigma$ level (one-sided; $p=5\times10^{-5}$).

\begin{figure}
    \centering
    \includegraphics[width=\linewidth]{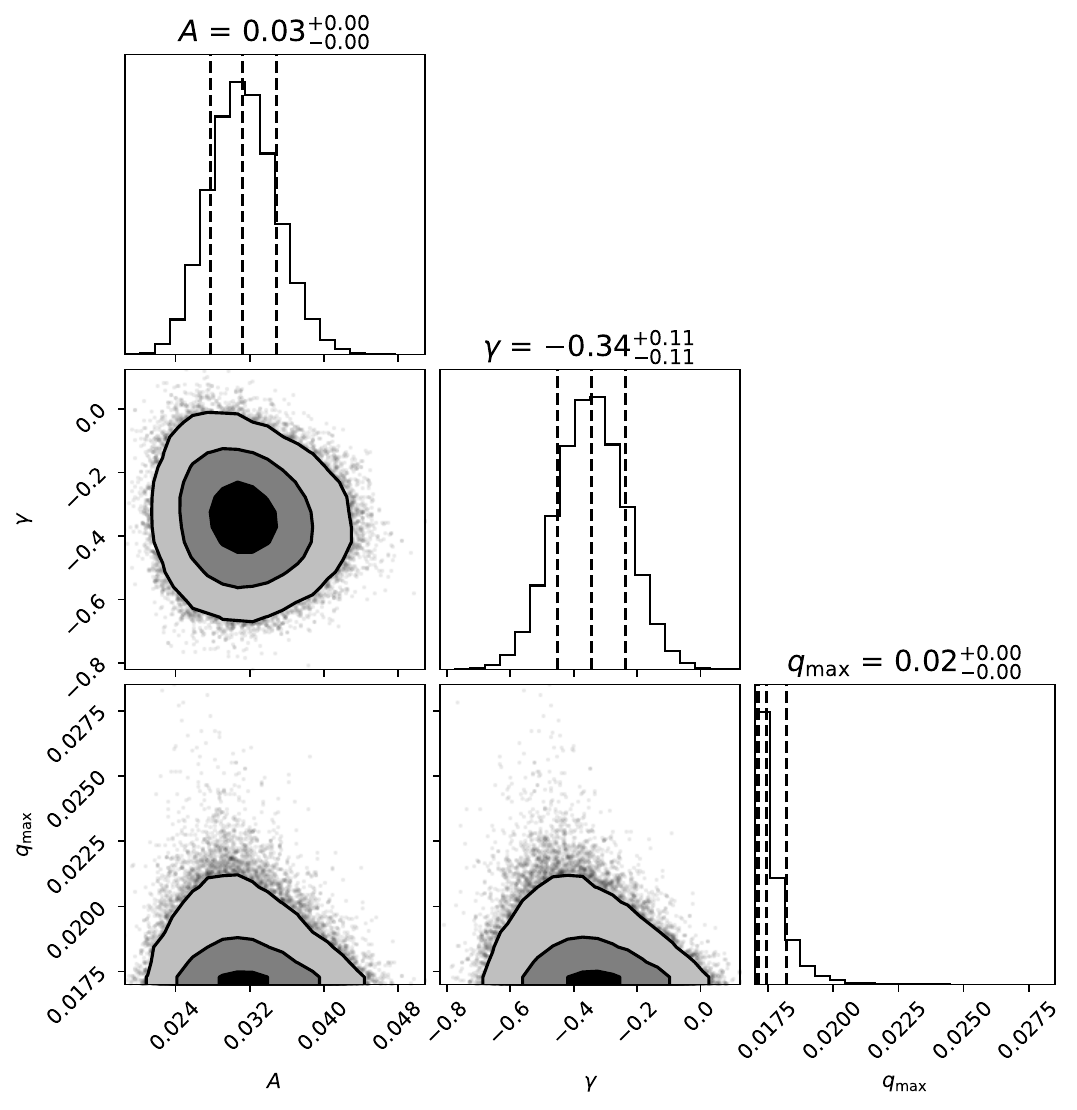}
    \caption{Corner plot illustrating the posterior distribution of the power-law mass-ratio function fitted to the combined KMTNet and MOA-II samples ($-3.4<\log(q)<-1.75$), with the upper truncation value $q_{\rm max}>0.017$ as a free parameter. Contours show the 1/2/3-$\sigma$ regions, with vertical lines in the 1D marginal plots showing the 16/50/84 percentiles. The parameter values differ from Equations 4--5, which adopted a fixed $q_{\rm max}=0.017$.}
\end{figure}

We then investigate whether there is a statistically significant truncation to this power-law mass-ratio distribution.
To this end, we assume that the power law distribution is truncated at some mass ratio larger than the largest observed value below the desert, namely $q=0.017$ (Table 1), and adopt a uniform prior $q_{\rm max}>0.017$ for this truncation value.
Based on planets detected over the full range of $-3.4<\log(q)<-1.75$, we derive truncations to the power law at $q_{\rm max}<0.0197$ and $q_{\rm max}<0.0224$ at the one-sided 2-$\sigma$ ($p=0.0228$) and 3-$\sigma$ ($p=0.00135$) significance (Figure 3).
As a robustness check, based on a power law fit restricted to the $q>0.002$ sample, we derive truncations to the power law at $q_{\rm max}<0.0202$ and $q_{\rm max}<0.0240$ at the same significance levels.
Thus, a 2-$\sigma$ truncation at $q=0.02$ is insensitive to the particular choice of the power-law fit.
These limits imply that none of the three detected desert events between the KMTNet and MOA-II samples originates from the same power-law distribution of gas-giant planets, which in turn, justifies our initial assumption of non-detection above $q>0.017$.

As a side note, if there were indeed a sub-Saturn desert below a fixed mass of approximately $M_{\rm min}=100M_\oplus$, then the increased giant-planet occurrence rates for higher mass hosts could be at least partially explained by the fact that higher mass hosts could form giant planets over a wider mass range.
By integrating the mass-ratio function (derived from the $q>0.002$ sample) from a host-mass-dependent $q_{\rm min}=M_{\rm min}/M_{\rm host}$ to a fixed upper limit of $q_{\rm max}=0.018$, the total occurrence rate should depend on the host mass as $M^{0.64\pm0.16}$, which is somewhat lower than the approximately linear dependency as suggested by radial velocity surveys \citep{johnson_giant_2010}.
This difference would imply that the normalization of the mass-ratio function may be dependent on the host mass, although the uncertainties from radial velocity surveys are quite large.

\section{Discussion}
We have identified a mass-ratio desert at $0.02\lesssim q\lesssim0.05$ around microlensing host stars, based on a statistically significant truncation of the microlensing companion mass-ratio distribution at $q=0.02$, and a pile-up of companions at $q\simeq0.05$.
Although the current sample is insufficient to quantify the dependency of the desert boundaries on the host mass, the fact that microlensing host masses span more than an order of magnitude (0.05--0.8 $M_\odot$), in contrast to the sharp truncation and aridness of the desert, suggests that this dependency cannot be large.
For example, one of the microlensing companions near the desert lower edge (OGLE-2017-BLG-1522Lb; $q=0.016$) is inferred to be a Jovian planet orbiting a brown-dwarf host \citep{jung_ogle-2017-blg-1522_2018}.

Furthermore, the desert boundaries are also consistent with those identified from direct imaging surveys around Sun-like and intermediate-mass stars \citep{duchene_low-mass_2022}, although direct imaging surveys are sensitive to companions at wider separations of 20--1000 au.
Therefore, our analysis based on microlensing data furthers the hypothesis that the canonical brown dwarf desert is fundamentally a feature in the mass ratio.

The persistence of this desert across stellar mass supports $q<0.02$ as a working definition of exoplanets.
Currently, the IAU working definition of an exoplanet includes an upper limit to both the physical mass and the mass ratio \citep{lecavelier_des_etangs_iau_2022}.
The maximum mass is based on the deuterium burning limit ($\sim$13 $M_{\rm J}$), whereas the limiting mass ratio of $q<0.04$ characterizes the hierarchical nature of planetary systems, based on the L4/L5 dynamical instability.
According to the IAU reference, some have proposed to relate the maximum mass of an exoplanet to the driest part of the brown dwarf desert, which were not favorably considered because the driest location is dependent on the host mass, nor is the brown dwarf desert completely dry.
The recognition of the brown dwarf desert as a mass-ratio desert alleviates both of these difficulties.

The use of $q<0.02$ as the defining threshold for exoplanets also relates to their distinct formation processes.
Previously, \cite{schlaufman_evidence_2018} observed that planets with $\lesssim4M_{J}$ orbit metal-rich solar-type stars, which is not the case for companions with $\gtrsim10M_{J}$.
It should be noted that these boundaries were identified using a clustering algorithm in the planet-mass and host-metallicity space, which is influenced by the planet mass distribution in their sample.
From their Figures 1 \& 2, we may see that the host metallicity of their sample does not trend significantly lower ([Fe/H]$<-$0.1) until $\gtrsim60M_{\rm J}$, consistent with the upper boundary of the mass-ratio desert.

This metallicity trend indicates that companions below the mass-ratio desert ($q<0.02$) are likely formed via core accretion and should be considered planets, whereas companions above the desert ($q>0.05$) are likely formed via gravitational instability.
The presence of a mass-ratio desert suggests that the formation of both populations of companions occurs in a scale-invariant fashion, with planetary companions growing to a maximum of approximately 2\% of their host masses.
Note that under this convention, the classification of companions as planets based on the mass ratio is independent of their classification as brown dwarfs based on the physical mass.

As for the non-planetary regime ($q>0.05$), \cite{han_brown_2022,han_probable_2023,han_brown_2023,han_microlensing_2024} have identified 13 brown dwarf candidates with $q<0.1$ across the 2016--2023 KMTNet seasons.
Apart from OGLE-2017-BLG-0614 ($q=0.05$; \citealt{han_brown_2022}), all other detections have mass ratios greater than $q>0.063$.
The absence of detections between $0.05<q<0.063$ may be suggestive of a break in the microlensing companion mass-ratio distribution around $q=0.06$--0.07 (before truncated at $q\simeq0.05$), but the evidence is quite tentative.
If real, this break could possibly reflect a transition between formation primarily via disk fragmentation below the break, and core fragmentation as binary stars above the break.

Notably, a similar break is already implicated in the work of \cite{duchene_low-mass_2022}, who fitted a power-law mass-ratio function to their detected low-mass companions to intermediate mass stars, with the lowest observed mass-ratio being $q\simeq0.1$. Similar to how we have derived a significant upper truncation to the giant-planet mass-ratio distribution, they obtained a lower truncation to their power-law distribution at $q_{\rm min}>0.073$ at the 3-$\sigma$ level. However, they eventually identified their desert upper edge as $q\simeq0.05$ from an abundance of directly imaged companions with $0.05<q<0.1$ in the literature. This population of $q>0.05$ companions below their derived power-law truncation is thus suggestive of a break to the companion mass-ratio distribution at $q=$ 0.07--0.1 for intermediate mass stars, at projected separations of $\sim100$ au.

Finally, we suggest that the detected microlensing companions in the mass-ratio desert likely orbit white dwarfs. As the typical 0.6$M_\odot$ white dwarf (e.g., \citealt{kilic_100_2020}) retains around one-third of its progenitor mass (e.g., \citealt{cunningham_initial-final_2024}), these companions will likely be initially below this desert around their main-sequence hosts, which would classify them as planets.
White-dwarf lenses are estimated to account for approximately 15\% of microlensing events \citep{koshimoto_bayesian_2020,lam_popsycle_2020}. If we assume that the population of gas giant planets orbiting white dwarfs follows the same power law distribution derived over $-3.4<\log(q)<-1.75$ but scaled down by a factor of 10 ($A\rightarrow A/10$ in Equation 1), then integrating it over the mass-ratio desert predicts $2.7^{+0.7}_{-0.6}$ expected detections between the KMTNet and MOA-II samples.
In other words, it is plausible that most microlensing companions detected in this desert orbit white dwarfs, which would imply that the mass-ratio desert around main-sequence stars is closer to being completely dry.

For context, direct imaging surveys of some nearby young stellar associations do not detect any companions over $0.01\lesssim q\lesssim 0.05$ and projected separations around 10 au, with both lower and higher mass ratios well populated \citep{gratton_stellar_2024}.
For the few directly imaged companions possibly in the mass-ratio desert (e.g., \citealt{hinkley_discovery_2015}), one should keep in mind that the inferred mass for directly imaged companions may be sensitive to the evolutionary model and assumed age.
In comparison, simultaneous fits to Hipparcos-GAIA astrometry and long-term radial-velocity data have placed at least some companions in the mass-ratio desert \citep{xiao_masses_2023,unger_exploring_2023}, whereas none in the sample of \cite{an_orbits_2025} within an semi-major axis of $\lesssim$ 20 au fall in this regime.
The latter work also noted the brown-dwarf desert to be slightly more pronounced in the mass ratio.
The upcoming GAIA data release 4 will be particularly useful for further investigations via this approach.

It could be challenging to determine whether microlensing hosts are indeed white dwarfs.
Previously, two microlensing systems have been inferred to have white dwarf primaries \citep{blackman_jovian_2021,zhang_earth-mass_2024}, based on the non-detection of the lens flux from high-angular-resolution imaging, under the assumption that they are main-sequence stars.
However, this approach would require the lens mass to be constrained in advance by high-order effects in the light-curve model, which is not the case for all three desert companions listed in Table 1.
For the present cases, a non-detection of the lens star --- after it has sufficiently separated from the source star --- can also admit brown dwarf or low-mass M dwarf interpretations, depending on the detection limit.

On the other hand, late-time detections of luminous microlensing hosts generally rule out stellar remnant interpretations, which enables one to construct a microlensing companion sample that excludes stellar remnant hosts.
The microlensing survey of the Roman Space Telescope is expected to detect more than 1000 exoplanets \citep{penny_predictions_2019}, with precise mass measurements based on lens flux measurements for approximately 40\% of all planet hosts \citep{terry_predictions_2025}.
Therefore, Roman will soon enable detailed investigations of this companion mass-ratio desert, including possible dependencies on the host properties.

\textit{Note added in proof.} After the final revision was submitted, \cite{giacalone_transition_2025} reported similar metallicity trends across the desert edges, whereas \cite{zandt_smooth_2025} found the brown-dwarf desert to extend to 10 au around Sun-like stars.
Moreover, \cite{zandt_smooth_2025} suggested a smooth transition between planets and brown dwarfs in the physical mass, which is compatible with our findings of an abrupt transition in the mass ratio.

\begin{acknowledgements}
I thank Clarissa Do \'O for helpful discussions.
I thank David Bennett and Daisuke Suzuki for clarifications on the MOA-II sample.
I thank Scott Gaudi, Quinn Konopacky, and Jean-Baptiste Ruffio for comments on the manuscript.
Support for this work was provided by NASA through the NASA Hubble Fellowship grant HST-HF2-51593 awarded by the Space Telescope Science Institute, which is operated by the
Association of Universities for Research in Astronomy, Inc., for NASA, under contract NAS5-26555.
I was supported by the Eric and Wendy Schmidt AI in Science Postdoctoral Fellowship, a Schmidt Futures program.
\end{acknowledgements}

\appendix
\restartappendixnumbering

\section{Overview of the KMTNet 2016--2019 Sample}

We have compiled the sample of 2016--2019 KMTNet unambiguous planets from the following works: 2016 Prime (Table 12; \citealt{shin_systematic_2023}), 2016 Sub-Prime (Table 16; \citealt{shin_systematic_2024}), 2017 Prime (Table 17; \citealt{ryu_systematic_2024}), 2017 Sub-Prime (Table 8; \citealt{gui_systematic_2024}), and 2017--2018 Prime and Sub-Prime (Table S3; \citealt{zang_microlensing_2025}).
The KMTNet systematic search for planetary events is conducted with the goal to ensure completeness for $q<0.03$ events. To this end, they modeled all possibly planetary events flagged by AnomalyFinder using their standard pipeline (i.e., online) photometry, and sent all events with $q<0.06$ solutions for re-reduction using a variant of the pySIS difference imaging algorithm \citep{albrow_difference_2009}.
As a general policy, the re-reduced data are then remodeled and published regardless of the final mass-ratio (e.g., \citealt{shin_systematic_2023,shin_systematic_2024}), although those failing the $q<0.03$ criteria are excluded from their final statistical analysis.
There exist some variations in this policy.
For example, for the 2018 prime field \citep{gould_systematic_2022}, the criterion for re-reduction was $q=0.05$, but all events with initial solutions of $q<0.06$ are being reported.
For the 2019 subprime field \citep{jung_systematic_2023}, the criteria for re-reduction is $q<0.06$, but only those with final mass ratios of $q<0.05$ are reported.

Among the 2016--2019 seasons, we identify a total of 5 events that passed the initial $q<0.06$ threshold but ended up with at least one possible ($\Delta\chi^2<10$) solution with $0.03<q<0.1$ based on the re-reduced data.
Three are from the 2019 season \citep{zang_systematic_2022}: KMT-2019-BLG-0814 ($q=0.05$), OGLE-2019-BLG-1067 ($q=0.047$), and KMT-2019-BLG-3301 ($q=0.053$). Three are from the 2016 season \citep{shin_systematic_2023}: KMT-2016-BLG-1222 ($q=0.088$ or $q=0.113$),
and OGLE-2016-BLG-0558 ($q=0.048\pm0.003$ or $q=0.057\pm0.003$). None were reported for the 2017 or 2018 seasons.

Recently, \cite{zang_microlensing_2025} conducted a statistical analysis of the $q<0.03$ planetary sample, and demonstrated a bimodal distribution of the mass-ratio function that peaks at super-Earth ($q\sim2\times10^{-5}$) and gas-giant ($q\sim2\times10^{-3}$) mass-ratios.
They demonstrated that a double-Gaussian mass-ratio distribution is significantly favored over a single power-law fit across $-6<\log(q)<-1.5$. The single power-law fit was shown to significantly over-predict the number of detections in three bins: $-6<\log(q)<-5$, $-3.6<\log(q)<-3$, and $-2<\log(q)<-1.5$.
The discrepancy for the final bin would be an initial indication of a possible desert, although this connection was not drawn in the previous work.
The sub-Saturn desert at $-3.6<\log(q)<-3$, initially and tentatively identified by \cite{yang_kmt-2016-blg-1836lb_2020}, was interpreted to be in agreement with the scenario of runaway gas accretion \citep{ida_toward_2004}.

\begin{table*}[ht]
  \centering
  \caption{Microlensing events in the extended MOA-II and KMTNet samples with at least one degenerate solution in the mass-ratio desert of $0.017<q<0.047$. The first row lists the best-fit light curve model, with $\Delta\chi^2$ indicating the relative goodness-of-fit for the alternative light-curve models. The designation of ``no preference'' indicates that the discovery paper does not strongly favor any of the degenerate models based on the light-curve goodness of fit, the presence of data systematics, or inferred properties of the source star. For KMT-2016-BLG-0212, the two degenerate solutions with $q\simeq0.005$ are merged due to their proximity.}
  \label{tab:microlensing_desert}
  \renewcommand{\arraystretch}{1.1}
  \begin{tabular}{lcc l}
    \hline
    Event                       & $q$ ($10^{-2}$)   & $s$            & Reference \\
    \hline
    KMT-2017-BLG-2197           & $1.85\pm0.29$     & $0.72\pm0.02$  & \cite{han_four_2025} \\
    ($\Delta\chi^2=0.2$)          & $2.36\pm0.35$     & $1.54\pm0.04$  &  \\
    \hline
    OGLE-2016-BLG-0613          & $2.90\pm0.20$     & $1.40\pm0.01$  & \cite{han_ogle-2016-blg-0613labb_2017} \\
                                (no preference)& $38.6\pm1.0$      & $0.74\pm0.01$  &  \\
                                (no preference)& $105\pm5$         & $1.94\pm0.02$  &  \\
    \hline
    KMT-2016-BLG-1107           & $3.60\pm0.50$     & $0.35\pm0.01$  & \cite{hwang_kmt-2016-blg-1107_2019} \\
                                (no preference) & $0.40\pm0.10$     & $2.97\pm0.10$  & \\
    \hline
    KMT-2016-BLG-0212           & $3.68\pm0.61$     & $0.83\pm0.01$  & \cite{hwang_kmt-2016-blg-0212_2018} \\
    ($\Delta\chi^2=6.6$)          & $0.005\pm0.001$   & $1.43\pm0.01$  &  \\
    ($\Delta\chi^2=8.7$)          & $0.008\pm0.002$   & $1.43\pm0.02$  &  \\
    \hline
  \end{tabular}
\end{table*}

\section{Ambiguous events in the KMTNet sample}
The criteria to qualify as an unambiguous planet to be included in the statistical planet sample is laid out in \cite{zang_microlensing_2025}.
To briefly summarize, the event shall not have a viable single-lens, binary-source (1L2S) light-curve model within $\Delta \chi^2<15$, nor any degenerate models ($\Delta \chi^2<10$) with a stellar-binary mass ratio ($q>0.03$).
If there are multiple planetary models within $\Delta \chi^2<10$, the difference in $\Delta\log q$ between any pair of degenerate solutions shall be within $\Delta\log q<0.25$.
For each planetary solution, the uncertainty in the mass ratio must be $\sigma(\log q)<0.2$.
Planets in binary-star systems are also excluded because stellar-binary events are excluded from the sample of events for the planet sensitivity calculation.

The rejected events due to degeneracies with a binary source (1L2S) model are accounted for in the detection efficiency and survey sensitivity calculations, but the excluded planets due to degeneracies in the mass ratio are not. According to \cite{zang_microlensing_2025}, this leads to a correction factor of 9.5\% to account for planets with uncertain mass ratios, and a maximum factor of 14.3\% if all events with alternative binary solutions are planets, which did not constitute a dominating error term for their analysis. In contrast, the ambiguous events over $0.017<q<0.047$ potentially dominate over the unambiguous events, making it important to consider them.

We identified events initially excluded due to degeneracies in the mass ratio, with at least one solution 1-$\sigma$ away from the desert boundaries (Table A1).
Regarding population statistics of the degenerate solutions, the multiplicity of M dwarfs is approximately $24\%$, with a probability density of $0.14$ per dex of $s$ at projected separations of a few au \citep{ward-duong_m-dwarfs_2015}.
The relative abundances between ice- and gas-giant planets are illustrated in Figure 2, and were already known from previous microlensing analysis (e.g., \citealt{suzuki_exoplanet_2016,zang_microlensing_2025}).
Therefore, ice giant planet and stellar binary solutions are favored over the desert solution by approximately two orders of magnitude, with gas giant solutions favored by one order of magnitude.

The light curve models formally favor the desert solution for OGLE-2016-BLG-0613 by $\Delta\chi^2\sim10$. However, the desert solution for OGLE-2016-BLG-0613 places the source star on an unlikely location on the color-magnitude diagram, leading the discovery paper to conclude that the desert solution is somewhat disfavored despite the preference by the light curve model.
Note that OGLE-2016-BLG-0613 was also excluded from the KMTNet statistical sample due to the binary host of the planet. However, the $q=0.03$ solution for this event relates to the mass ratio between the binary hosts, not the planet to the lens mass, so this exclusion criterion does not apply to the present analysis.
As for KMT-2016-BLG-1107, the discovery paper concluded the apparent $\Delta\chi^2=42$ preference for the $q=0.036$ solution is an artifact due to large data systematics. Therefore, we consider the desert and planetary solutions equally likely.
Due to the strong prior preference for non-desert solutions, we reject the desert solutions for KMT-2016-BLG-1107 and OGLE-2016-BLG-0613, for which the observed data do not have a strong preference among the degenerate models.

For KMT-2016-BLG-0212, the desert light curve model is formally favored by $\Delta\chi^2=6.6$--8.7 compared to each of the three ice-giant solutions, implying a likelihood ratio of 27--77:1.
It is important to note that microlensing photometry typically does not strictly follow Gaussian statistics, so differences in $\chi^2$ should not be interpreted at face value as significance measures.
However, even if this likelihood ratio is taken at face value, it does not overcome the prior preference for any one ice giant solution by approximately two orders of magnitude. 
As there exists a total of three ice giant solutions that are at least mildly favored over the desert solution, we conclude that the desert solution remains strongly ($\sim$1:10) disfavored over the totality of the three ice-giant planet models combined.

The remaining ambiguous event is KMT-2017-BLG-2197, which has one solution at the desert boundary and one slightly above it. For the two solutions combined with equal weights ($q=0.021\pm0.004$), the likelihood ratio of the mass ratio being below or above a nominal $q=0.018$ desert edge is approximately 1:3, which compares to the order of magnitude drop in the occurrence rate density across this edge.
Therefore, Bayesian statistics favor KMT-2017-BLG-2197 to be a non-desert event, although less so compared to the three other ambiguous events.
Its mass-ratio uncertainty remains large even with a population prior, and will not be informative of either the desert edge or the desert abundance. As such, we exclude it from our statistical analysis.


\begin{thebibliography}{}
\expandafter\ifx\csname natexlab\endcsname\relax\def\natexlab#1{#1}\fi
\providecommand{\url}[1]{\href{#1}{#1}}
\providecommand{\dodoi}[1]{doi:~\href{http://doi.org/#1}{\nolinkurl{#1}}}
\providecommand{\doeprint}[1]{\href{http://ascl.net/#1}{\nolinkurl{http://ascl.net/#1}}}
\providecommand{\doarXiv}[1]{\href{https://arxiv.org/abs/#1}{\nolinkurl{https://arxiv.org/abs/#1}}}

\bibitem[{Albrow {et~al.}(2009)Albrow, Horne, Bramich, Fouqué, Miller, Beaulieu, Coutures, Menzies, Williams, Batista, Bennett, Brillant, Cassan, Dieters, Prester, Donatowicz, Greenhill, Kains, Kane, Kubas, Marquette, Pollard, Sahu, Tsapras, Wambsganss, \& Zub}]{albrow_difference_2009}
Albrow, M.~D., Horne, K., Bramich, D.~M., {et~al.} 2009, Monthly Notices of the Royal Astronomical Society, 397, 2099, \dodoi{10.1111/j.1365-2966.2009.15098.x}

\bibitem[{An {et~al.}(2025)An, Brandt, Brandt, \& Venner}]{an_orbits_2025}
An, Q., Brandt, T.~D., Brandt, G.~M., \& Venner, A. 2025, The Astrophysical Journal Supplement Series, 280, 61, \dodoi{10.3847/1538-4365/adfa99}

\bibitem[{Blackman {et~al.}(2021)Blackman, Beaulieu, Bennett, Danielski, Alard, Cole, Vandorou, Ranc, Terry, Bhattacharya, Bond, Bachelet, Veras, Koshimoto, Batista, \& Marquette}]{blackman_jovian_2021}
Blackman, J.~W., Beaulieu, J.~P., Bennett, D.~P., {et~al.} 2021, Nature, 598, 272, \dodoi{10.1038/s41586-021-03869-6}

\bibitem[{Bond {et~al.}(2004)Bond, Udalski, Jaroszyński, Rattenbury, Paczyński, Soszyński, Wyrzykowski, Szymański, Kubiak, Szewczyk, Żebruń, Pietrzyński, Abe, Bennett, Eguchi, Furuta, Hearnshaw, Kamiya, Kilmartin, Kurata, Masuda, Matsubara, Muraki, Noda, Okajima, Sako, Sekiguchi, Sullivan, Sumi, Tristram, Yanagisawa, Yock, \& Collaborations}]{bond_ogle_2004}
Bond, I.~A., Udalski, A., Jaroszyński, M., {et~al.} 2004, The Astrophysical Journal, 606, L155, \dodoi{10.1086/420928}

\bibitem[{Cunningham {et~al.}(2024)Cunningham, Tremblay, \& W. O’Brien}]{cunningham_initial-final_2024}
Cunningham, T., Tremblay, P.-E., \& W. O’Brien, M. 2024, Monthly Notices of the Royal Astronomical Society, 527, 3602, \dodoi{10.1093/mnras/stad3275}

\bibitem[{Duch{\^e}ne {et~al.}(2022)Duch{\^e}ne, Oon, De Rosa, Kantorski, Coy, Wang, Thomas, Patience, Pueyo, Nielsen, \& Konopacky}]{duchene_low-mass_2022}
Duch{\^e}ne, G., Oon, J.~T., De Rosa, R.~J., {et~al.} 2022, Monthly Notices of the Royal Astronomical Society, 519, 778, \dodoi{10.1093/mnras/stac3527}

\bibitem[{Foreman-Mackey {et~al.}(2013)Foreman-Mackey, Hogg, Lang, \& Goodman}]{foreman-mackey_emcee_2013}
Foreman-Mackey, D., Hogg, D.~W., Lang, D., \& Goodman, J. 2013, Publications of the Astronomical Society of the Pacific, 125, 306, \dodoi{10.1086/670067}

\bibitem[{Giacalone {et~al.}(2025)Giacalone, Howard, Gilbert, Zandt, Petigura, \& Handley}]{giacalone_transition_2025}
Giacalone, S., Howard, A.~W., Gilbert, G.~J., {et~al.} 2025, The {Transition} from {Giant} {Planets} to {Brown} {Dwarfs} beyond 1 au from the {Stellar} {Metallicity} {Distribution},  arXiv, \dodoi{10.48550/arXiv.2511.11818}

\bibitem[{Gould {et~al.}(2022)Gould, Han, Zang, Yang, Hwang, Udalski, Bond, Albrow, Chung, Jung, Ryu, Shin, Shvartzvald, Yee, Cha, Kim, Kim, Kim, Lee, Lee, Lee, Park, Pogge, Mróz, Szymański, Skowron, Poleski, Soszyński, Pietrukowicz, Kozłowski, Ulaczyk, Rybicki, Iwanek, Wrona, Abe, Barry, Bennett, Bhattacharya, Fujii, Fukui, Hirao, Silva, Kirikawa, Kondo, Koshimoto, Matsubara, Matsumoto, Miyazaki, Muraki, Okamura, Olmschenk, Ranc, Rattenbury, Satoh, Sumi, Suzuki, Toda, Tristram, Vandorou, Yama, Beichman, Bryden, Novati, Gaudi, Henderson, Penny, Jacklin, \& Stassun}]{gould_systematic_2022}
Gould, A., Han, C., Zang, W., {et~al.} 2022, Astronomy \& Astrophysics, 664, A13, \dodoi{10.1051/0004-6361/202243744}

\bibitem[{Gratton {et~al.}(2024)Gratton, Bonavita, Mesa, Desidera, Zurlo, Marino, D’Orazi, Rigliaco, Nascimbeni, Barbato, Columba, \& Squicciarini}]{gratton_stellar_2024}
Gratton, R., Bonavita, M., Mesa, D., {et~al.} 2024, Astronomy \& Astrophysics, 685, A119, \dodoi{10.1051/0004-6361/202348393}

\bibitem[{Grether \& Lineweaver(2006)}]{grether_how_2006}
Grether, D., \& Lineweaver, C.~H. 2006, The Astrophysical Journal, 640, 1051, \dodoi{10.1086/500161}

\bibitem[{Gui {et~al.}(2024)Gui, Zang, Zhai, Ryu, Udalski, Yang, Han, Mao, {Leading Authors}, Albrow, Chung, Gould, Hwang, Jung, Shin, Shvartzvald, Yee, Cha, Kim, Kim, Kim, Lee, Lee, Lee, Park, Pogge, {The KMTNet Collaboration}, Mróz, Szymański, Skowron, Poleski, Soszyński, Pietrukowicz, Kozłowski, Ulaczyk, Rybicki, Iwanek, Wrona, Gromadzki, {The OGLE Collaboration}, Wang, Zhang, Kuang, Qian, Zhu, \& {The MAP Collaboration}}]{gui_systematic_2024}
Gui, Y., Zang, W., Zhai, R., {et~al.} 2024, The Astronomical Journal, 168, 49, \dodoi{10.3847/1538-3881/ad4ce5}

\bibitem[{Han {et~al.}(2017{\natexlab{a}})Han, Udalski, Gould, Bond, {and}, Albrow, Chung, Jung, Ryu, Shin, Yee, Zhu, Cha, Kim, Kim, Lee, Lee, Park, {KMTNet Collaboration}, Skowron, Mróz, Pietrukowicz, Kozłowski, Poleski, Szymański, Soszyński, Ulaczyk, Pawlak, {OGLE Collaboration}, Abe, Asakura, Barry, Bennett, Bhattacharya, Donachie, Evans, Fukui, Hirao, Itow, Koshimoto, Li, Ling, Masuda, Matsubara, Muraki, Nagakane, Ohnishi, Ranc, Rattenbury, Saito, Sharan, Sullivan, Sumi, Suzuki, Tristram, Yamada, Yamada, Yonehara, \& {MOA Collaboration}}]{han_ogle-2016-blg-0263lb_2017}
Han, C., Udalski, A., Gould, A., {et~al.} 2017{\natexlab{a}}, The Astronomical Journal, 154, 133, \dodoi{10.3847/1538-3881/aa859a}

\bibitem[{Han {et~al.}(2017{\natexlab{b}})Han, Udalski, A, Lee, Shvartzvald, Zang, Mao, Kozłowski, Albrow, Chung, Hwang, Jung, Kim, Kim, Ryu, Shin, Yee, Zhu, Cha, Kim, Kim, Lee, Park, {(The KMTNet Collaboration)}, Skowron, Mróz, Pietrukowicz, Poleski, Szymański, Soszyński, Ulaczyk, Pawlak, {(The OGLE Collaboration)}, Beichman, Bryden, Novati, Gaudi, Henderson, Howell, Jacklin, {(The UKIRT Microlensing Team)}, Penny, Fouqué, Wang, \& {(CFHT-K2C9 Microlensing Collaboration)}}]{han_ogle-2016-blg-0613labb_2017}
Han, C., Udalski, A., A, G., {et~al.} 2017{\natexlab{b}}, The Astronomical Journal, 154, 223, \dodoi{10.3847/1538-3881/aa9179}

\bibitem[{Han {et~al.}(2021)Han, Udalski, Kim, Lee, Albrow, Chung, Gould, Hwang, Jung, Ryu, Shin, Shvartzvald, Yee, Zang, Cha, Kim, Kim, Kim, Lee, Lee, Park, Pogge, Mróz, Szymański, Skowron, Poleski, Soszyński, Pietrukowicz, Kozłowski, Ulaczyk, Rybicki, Iwanek, Wrona, \& Gromadzki}]{han_three_2021}
Han, C., Udalski, A., Kim, D., {et~al.} 2021, Astronomy and Astrophysics, 650, A89, \dodoi{10.1051/0004-6361/202140758}

\bibitem[{Han {et~al.}(2022)Han, Ryu, Shin, Jung, Kim, Hirao, Bozza, Albrow, Zang, Udalski, Bond, Chung, Gould, Hwang, Shvartzvald, Yang, Cha, Kim, Kim, Kim, Lee, Lee, Yee, Lee, Park, Pogge, Mróz, Szymański, Skowron, Poleski, Soszyński, Pietrukowicz, Kozłowski, Ulaczyk, Rybicki, Iwanek, Wrona, Abe, Barry, Bennett, Bhattacharya, Fujii, Fukui, Silva, Kirikawa, Kondo, Koshimoto, Matsubara, Matsumoto, Miyazaki, Muraki, Okamura, Olmschenk, Ranc, Rattenbury, Satoh, Sumi, Suzuki, Toda, Tristram, Vandorou, Yama, \& Itow}]{han_brown_2022}
Han, C., Ryu, Y.-H., Shin, I.-G., {et~al.} 2022, Astronomy \& Astrophysics, 667, A64, \dodoi{10.1051/0004-6361/202244186}

\bibitem[{Han {et~al.}(2023{\natexlab{a}})Han, Kil~Jung, Kim, Gould, Bozza, Bond, Chung, Albrow, Hwang, Ryu, Shin, Shvartzvald, Yang, Zang, Cha, Kim, Kim, Kim, Lee, Lee, Yee, Lee, Park, Pogge, Abe, Barry, Bennett, Bhattacharya, Fujii, Fukui, Hirao, Ishitani~Silva, Kirikawa, Kondo, Koshimoto, Matsubara, Matsumoto, Miyazaki, Muraki, Okamura, Olmschenk, Ranc, Rattenbury, Satoh, Sumi, Suzuki, Toda, Tristram, Vandorou, Yama, \& Itow}]{han_probable_2023}
Han, C., Kil~Jung, Y., Kim, D., {et~al.} 2023{\natexlab{a}}, Astronomy and Astrophysics, 675, A71, \dodoi{10.1051/0004-6361/202245455}

\bibitem[{Han {et~al.}(2023{\natexlab{b}})Han, Jung, Bond, Chung, Albrow, Gould, Hwang, Lee, Ryu, Shin, Shvartzvald, Yang, Yee, Zang, Cha, Kim, Kim, Kim, Lee, Lee, Park, Pogge, Abe, Barry, Bennett, Bhattacharya, Fujii, Fukui, Hamada, Hirao, Silva, Itow, Kirikawa, Koshimoto, Matsubara, Miyazaki, Muraki, Olmschenk, Ranc, Rattenbury, Satoh, Sumi, Suzuki, Tomoyoshi, Tristram, Vandorou, Yama, \& Yamashita}]{han_brown_2023}
Han, C., Jung, Y.~K., Bond, I.~A., {et~al.} 2023{\natexlab{b}}, Astronomy \& Astrophysics, 678, A190, \dodoi{10.1051/0004-6361/202347014}

\bibitem[{Han {et~al.}(2024)Han, Bond, Udalski, Lee, Gould, Albrow, Chung, Hwang, Jung, Ryu, Shvartzvald, Shin, Yee, Yang, Zang, Cha, Kim, Kim, Kim, Lee, Lee, Park, Pogge, Abe, Bando, Barry, Bennett, Bhattacharya, Fujii, Fukui, Hamada, Hamada, Hamasaki, Hirao, Silva, Itow, Kirikawa, Koshimoto, Matsubara, Miyazaki, Muraki, Nagai, Nunota, Olmschenk, Ranc, Rattenbury, Satoh, Sumi, Suzuki, Tomoyoshi, Tristram, Vandorou, Yama, Yamashita, Mróz, Szymański, Skowron, Poleski, Soszyński, Pietrukowicz, Kozłowski, Rybicki, Iwanek, Ulaczyk, Wrona, Gromadzki, \& Mróz}]{han_microlensing_2024}
Han, C., Bond, I.~A., Udalski, A., {et~al.} 2024, Astronomy \& Astrophysics, 691, A237, \dodoi{10.1051/0004-6361/202451416}

\bibitem[{Han {et~al.}(2025)Han, Albrow, Lee, Chung, Gould, Hwang, Jung, Ryu, Shvartzvald, Shin, Yee, Yang, Zang, Cha, Kim, Kim, Kim, Lee, Lee, Park, \& Pogge}]{han_four_2025}
Han, C., Albrow, M.~D., Lee, C.-U., {et~al.} 2025, The Astronomical Journal, 169, 288, \dodoi{10.3847/1538-3881/adc5e7}

\bibitem[{Hinkley {et~al.}(2015)Hinkley, Kraus, Ireland, Cheetham, Carpenter, Tuthill, Lacour, Evans, \& Haubois}]{hinkley_discovery_2015}
Hinkley, S., Kraus, A.~L., Ireland, M.~J., {et~al.} 2015, The Astrophysical Journal Letters, 806, L9, \dodoi{10.1088/2041-8205/806/1/L9}

\bibitem[{Hwang {et~al.}(2018)Hwang, Kim, Kim, Gould, Albrow, Chung, Han, Jung, Ryu, Shin, Shvartzvald, Yee, Zang, Zhu, Cha, Kim, Lee, Lee, Lee, Park, \& Pogge}]{hwang_kmt-2016-blg-0212_2018}
Hwang, K.~H., Kim, H.~W., Kim, D.~J., {et~al.} 2018, Journal of Korean Astronomical Society, 51, 197, \dodoi{10.5303/JKAS.2018.51.6.197}

\bibitem[{Hwang {et~al.}(2019)Hwang, Ryu, Kim, Albrow, Chung, Gould, Han, Jung, Shin, Shvartzvald, Yee, Zang, Cha, Kim, Kim, Lee, Lee, Lee, Park, \& Pogge}]{hwang_kmt-2016-blg-1107_2019}
Hwang, K.-H., Ryu, Y.-H., Kim, H.-W., {et~al.} 2019, The Astronomical Journal, 157, 23, \dodoi{10.3847/1538-3881/aaf16e}

\bibitem[{Ida \& Lin(2004)}]{ida_toward_2004}
Ida, S., \& Lin, D. N.~C. 2004, The Astrophysical Journal, 604, 388, \dodoi{10.1086/381724}

\bibitem[{Johnson {et~al.}(2010)Johnson, Aller, Howard, \& Crepp}]{johnson_giant_2010}
Johnson, J.~A., Aller, K.~M., Howard, A.~W., \& Crepp, J.~R. 2010, Publications of the Astronomical Society of the Pacific, 122, 905, \dodoi{10.1086/655775}

\bibitem[{Jung {et~al.}(2018)Jung, Udalski, Gould, Ryu, Yee, {and}, Han, Albrow, Lee, Kim, Hwang, Chung, Shin, Zhu, Cha, Kim, Lee, Park, Lee, Kim, Pogge, Collaboration), Szymański, Mróz, Poleski, Skowron, Pietrukowicz, Soszyński, Kozłowski, Ulaczyk, Pawlak, Rybicki, \& Collaboration)}]{jung_ogle-2017-blg-1522_2018}
Jung, Y.~K., Udalski, A., Gould, A., {et~al.} 2018, The Astronomical Journal, 155, 219, \dodoi{10.3847/1538-3881/aabb51}

\bibitem[{Jung {et~al.}(2023)Jung, Zang, Wang, Han, Gould, Udalski, Albrow, Chung, Hwang, Ryu, Shin, Shvartzvald, Yang, Yee, Cha, Kim, Kim, Lee, Lee, Lee, Park, Pogge, {KMTNet Collaboration}, Szymański, Skowron, Poleski, Soszyński, Pietrukowicz, Kozłowski, Ulaczyk, Rybicki, Iwanek, Wrona, {OGLE Collaboration}, Green, Hennerley, Marmont, Mao, Maoz, McCormick, Natusch, Penny, Porritt, Zhu, {Tsinghua Team}, \& {FUN Follow-Up Team}}]{jung_systematic_2023}
Jung, Y.~K., Zang, W., Wang, H., {et~al.} 2023, The Astronomical Journal, 165, 226, \dodoi{10.3847/1538-3881/accb8f}

\bibitem[{Kilic {et~al.}(2020)Kilic, Bergeron, Kosakowski, Brown, Agüeros, \& Blouin}]{kilic_100_2020}
Kilic, M., Bergeron, P., Kosakowski, A., {et~al.} 2020, The Astrophysical Journal, 898, 84, \dodoi{10.3847/1538-4357/ab9b8d}

\bibitem[{Koshimoto {et~al.}(2020)Koshimoto, Bennett, \& Suzuki}]{koshimoto_bayesian_2020}
Koshimoto, N., Bennett, D.~P., \& Suzuki, D. 2020, The Astronomical Journal, 159, 268, \dodoi{10.3847/1538-3881/ab8adf}

\bibitem[{Lam {et~al.}(2020)Lam, Lu, Hosek, Dawson, \& Golovich}]{lam_popsycle_2020}
Lam, C.~Y., Lu, J.~R., Hosek, M.~W., Dawson, W.~A., \& Golovich, N.~R. 2020, The Astrophysical Journal, 889, 31, \dodoi{10.3847/1538-4357/ab5fd3}

\bibitem[{Lecavelier~des Etangs \& Lissauer(2022)}]{lecavelier_des_etangs_iau_2022}
Lecavelier~des Etangs, A., \& Lissauer, J.~J. 2022, New Astronomy Reviews, 94, 101641, \dodoi{10.1016/j.newar.2022.101641}

\bibitem[{Marcy {et~al.}(1998)Marcy, Butler, Vogt, Fischer, \& Lissauer}]{marcy_planetary_1998}
Marcy, G.~W., Butler, R.~P., Vogt, S.~S., Fischer, D., \& Lissauer, J.~J. 1998, The Astrophysical Journal, 505, L147, \dodoi{10.1086/311623}

\bibitem[{Mignon {et~al.}(2025)Mignon, Delfosse, Meunier, Chaverot, Burn, Bonfils, Bouchy, Astudillo-Defru, Curto, Gaisne, Udry, Forveille, Segransan, Lovis, Santos, \& Mayor}]{mignon_radial_2025}
Mignon, L., Delfosse, X., Meunier, N., {et~al.} 2025, Astronomy \& Astrophysics, 700, A146, \dodoi{10.1051/0004-6361/202451142}

\bibitem[{Penny {et~al.}(2019)Penny, Scott~Gaudi, Kerins, Rattenbury, Mao, Robin, \& Calchi~Novati}]{penny_predictions_2019}
Penny, M.~T., Scott~Gaudi, B., Kerins, E., {et~al.} 2019, The Astrophysical Journal Supplement Series, 241, 3, \dodoi{10.3847/1538-4365/aafb69}

\bibitem[{Poleski {et~al.}(2017)Poleski, Udalski, Bond, Beaulieu, Clanton, Gaudi, Szymański, Soszyński, Pietrukowicz, Kozłowski, Skowron, Wyrzykowski, Ulaczyk, Bennett, Sumi, Suzuki, Rattenbury, Koshimoto, Abe, Asakura, Barry, Bhattacharya, Donachie, Evans, Fukui, Hirao, Itow, Li, Ling, Masuda, Matsubara, Muraki, Nagakane, Ohnishi, Ranc, Saito, Sharan, Sullivan, Tristram, Yamada, Yamada, Yonehara, Batista, \& Marquette}]{poleski_companion_2017}
Poleski, R., Udalski, A., Bond, I.~A., {et~al.} 2017, Astronomy \& Astrophysics, 604, A103, \dodoi{10.1051/0004-6361/201730928}

\bibitem[{Ranc {et~al.}(2015)Ranc, Cassan, Albrow, Kubas, Bond, Batista, Beaulieu, Bennett, Dominik, Dong, Fouqué, Gould, Greenhill, Jørgensen, Kains, Menzies, Sumi, Bachelet, Coutures, Dieters, Dominis~Prester, Donatowicz, Gaudi, Han, Hundertmark, Horne, Kane, Lee, Marquette, Park, Pollard, Sahu, Street, Tsapras, Wambsganss, Williams, Zub, Abe, Fukui, Itow, Masuda, Matsubara, Muraki, Ohnishi, Rattenbury, Saito, Sullivan, Sweatman, Tristram, Yock, \& Yonehara}]{ranc_moa-2007-blg-197_2015}
Ranc, C., Cassan, A., Albrow, M.~D., {et~al.} 2015, Astronomy \& Astrophysics, 580, A125, \dodoi{10.1051/0004-6361/201525791}

\bibitem[{Ryu {et~al.}(2024)Ryu, Udalski, Yee, Zang, Shvartzvald, Han, Gould, Albrow, Chung, Hwang, Jung, Shin, Yang, Cha, Kim, Kim, Lee, Lee, Lee, Park, Pogge, Wang, {KMTNet Collaboration}, Mróz, Szymański, Skowron, Poleski, Soszyński, Pietrukowicz, Kozłowski, Ulaczyk, Rybicki, Iwanek, Wrona, {OGLE Collaboration}, Beichman, Bryden, Carey, Henderson, Calchi~Novati, Zhu, {SPITZER Team}, Jacklin, Penny, \& {UKIRT Team}}]{ryu_systematic_2024}
Ryu, Y.-H., Udalski, A., Yee, J.~C., {et~al.} 2024, The Astronomical Journal, 167, 88, \dodoi{10.3847/1538-3881/ad1888}

\bibitem[{Schlaufman(2018)}]{schlaufman_evidence_2018}
Schlaufman, K.~C. 2018, The Astrophysical Journal, 853, 37, \dodoi{10.3847/1538-4357/aa961c}

\bibitem[{Shin {et~al.}(2023)Shin, Yee, Zang, Yang, Hwang, Han, Gould, Udalski, Bond, {Leading authors}, Albrow, Chung, Jung, Ryu, Shvartzvald, Cha, Kim, Kim, Lee, Lee, Lee, Park, Pogge, {The KMTNet Collaboration}, Mróz, Szymański, Skowron, Poleski, Soszyński, Pietrukowicz, Kozłowski, Rybicki, Iwanek, Ulaczyk, Wrona, Gromadzki, {The OGLE Collaboration}, Abe, Barry, Bennett, Bhattacharya, Fujii, Fukui, Hamada, Hirao, Silva, Itow, Kirikawa, Kondo, Koshimoto, Matsubara, Miyazaki, Muraki, Olmschenk, Ranc, Rattenbury, Satoh, Sumi, Suzuki, Tomoyoshi, Tristram, Vandorou, Yama, Yamashita, \& {the MOA Collaboration}}]{shin_systematic_2023}
Shin, I.-G., Yee, J.~C., Zang, W., {et~al.} 2023, The Astronomical Journal, 166, 104, \dodoi{10.3847/1538-3881/ace96d}

\bibitem[{Shin {et~al.}(2024)Shin, Yee, Zang, Han, Yang, Gould, Lee, Udalski, Sumi, Albrow, Chung, Hwang, Jung, Ryu, Shvartzvald, Cha, Kim, Kim, Kim, Lee, Lee, Park, Pogge, Mróz, Szymański, Skowron, Poleski, Soszyński, Pietrukowicz, Kozłowski, Rybicki, Iwanek, Ulaczyk, Wrona, Gromadzki, Abe, Bando, Barry, Bennett, Bhattacharya, Bond, Fujii, Fukui, Hamada, Hamada, Hamasaki, Hirao, Ishitani~Silva, Itow, Kirikawa, Koshimoto, Matsubara, Miyazaki, Muraki, Nagai, Nunota, Olmschenk, Ranc, Rattenbury, Satoh, Suzuki, Tomoyoshi, Tristram, Vandorou, Yama, \& Yamashita}]{shin_systematic_2024}
---. 2024, Systematic {KMTNet} {Planetary} {Anomaly} {Search}. {XI}. {Complete} {Sample} of 2016 {Sub}-{Prime} {Field} {Planets}, \dodoi{10.48550/arXiv.2401.04256}

\bibitem[{Shvartzvald {et~al.}(2014)Shvartzvald, Maoz, Kaspi, Sumi, Udalski, Gould, Bennett, Han, Abe, Bond, Botzler, Freeman, Fukui, Fukunaga, Itow, Koshimoto, Ling, Masuda, Matsubara, Muraki, Namba, Ohnishi, Rattenbury, Saito, Sullivan, Sweatman, Suzuki, Tristram, Wada, Yock, Skowron, Kozłowski, Szymański, Kubiak, Pietrzyński, Soszyński, Ulaczyk, Wyrzykowski, Poleski, \& Pietrukowicz}]{shvartzvald_moa-2011-blg-322lb_2014}
Shvartzvald, Y., Maoz, D., Kaspi, S., {et~al.} 2014, Monthly Notices of the Royal Astronomical Society, 439, 604, \dodoi{10.1093/mnras/stt2477}

\bibitem[{Shvartzvald {et~al.}(2016)Shvartzvald, Maoz, Udalski, Sumi, Friedmann, Kaspi, Poleski, Szymański, Skowron, Kozłowski, Wyrzykowski, Mróz, Pietrukowicz, Pietrzyński, Soszyński, Ulaczyk, Abe, Barry, Bennett, Bhattacharya, Bond, Freeman, Inayama, Itow, Koshimoto, Ling, Masuda, Fukui, Matsubara, Muraki, Ohnishi, Rattenbury, Saito, Sullivan, Suzuki, Tristram, Wakiyama, \& Yonehara}]{shvartzvald_frequency_2016}
Shvartzvald, Y., Maoz, D., Udalski, A., {et~al.} 2016, Monthly Notices of the Royal Astronomical Society, 457, 4089, \dodoi{10.1093/mnras/stw191}

\bibitem[{Street {et~al.}(2013)Street, Choi, Tsapras, Han, Furusawa, Hundertmark, Gould, Sumi, Bond, Wouters, Zellem, Udalski, {The RoboNet Collaboration}, Snodgrass, Horne, Dominik, Browne, Kains, Bramich, Bajek, Steele, Ipatov, {The MOA Collaboration}, Abe, Bennett, Botzler, Chote, Freeman, Fukui, Harris, Itow, Ling, Masuda, Matsubara, Miyake, Muraki, Nagayama, Nishimaya, Ohnishi, Rattenbury, Saito, Sullivan, Suzuki, Sweatman, Tristram, Wada, Yock, {The OGLE Collaboration}, Szymański, Kubiak, Pietrzyński, Soszyński, Poleski, Ulaczyk, Wyrzykowski, {The μFUN Collaboration}, Yee, Dong, Shin, Lee, Skowron, Andrade De~Almeida, DePoy, Gaudi, Hung, Jablonski, Kaspi, Klein, Hwang, Koo, Maoz, Muñoz, Pogge, Polishhook, Shporer, McCormick, Christie, Natusch, Allen, Drummond, Moorhouse, Thornley, Knowler, Bos, Bolt, {The PLANET Collaboration}, Beaulieu, Albrow, Batista, Brillant, Caldwell, Cassan, Cole, Corrales, Coutures, Dieters, Dominis~Prester, Donatowicz, Fouqué, Bachelet, Greenhill, Kane, Kubas, Marquette,
  Martin, Menzies, Pollard, Sahu, Wambsganss, Williams, Zub, {MiNDSTEp}, Alsubai, Bozza, Burgdorf, Calchi~Novati, Dodds, Dreizler, Finet, Gerner, Hardis, Harpsøe, Hessman, Hinse, Jørgensen, Kerins, Liebig, Mancini, Mathiasen, Penny, Proft, Rahvar, Ricci, Scarpetta, Schäfer, Schönebeck, Southworth, \& Surdej}]{street_moa-2010-blg-073l_2013}
Street, R.~A., Choi, J.-Y., Tsapras, Y., {et~al.} 2013, The Astrophysical Journal, 763, 67, \dodoi{10.1088/0004-637X/763/1/67}

\bibitem[{Suzuki {et~al.}(2016)Suzuki, Bennett, Sumi, Bond, Rogers, Abe, Asakura, Bhattacharya, Donachie, Freeman, Fukui, Hirao, Itow, Koshimoto, Li, Ling, Masuda, Matsubara, Muraki, Nagakane, Onishi, Oyokawa, Rattenbury, Saito, Sharan, Shibai, Sullivan, Tristram, \& and}]{suzuki_exoplanet_2016}
Suzuki, D., Bennett, D.~P., Sumi, T., {et~al.} 2016, The Astrophysical Journal, 833, 145, \dodoi{10.3847/1538-4357/833/2/145}

\bibitem[{Terry {et~al.}(2025)Terry, Bachelet, Zohrabi, Verma, Crisp, Huston, McGee, Penny, Abrams, Albrow, Anderson, Bagheri, Beaulieu, Bellini, Bennett, Bergsten, Bhadra, Bhattacharya, Bond, Bozza, Brandon, Novati, Carey, Christiansen, DeRocco, Gaudi, Hulberg, Silva, Jones, Kerins, Khakpash, Kruszynska, Lam, Lu, Malpas, Miyazaki, Mroz, Murlidhar, Nataf, Newman, Olmschenk, Poleski, Ranc, Rattenbury, Rybicki, Saggese, Sobeck, Stassun, Stephan, Street, Sumi, Suzuki, Vandorou, Vyas, Yee, Zang, \& Zhang}]{terry_predictions_2025}
Terry, S.~K., Bachelet, E., Zohrabi, F., {et~al.} 2025, Predictions of the {Nancy} {Grace} {Roman} {Space} {Telescope} {Galactic} {Exoplanet} {Survey}. {IV}. {Lens} {Mass} and {Distance} {Measurements},  arXiv, \dodoi{10.48550/arXiv.2510.13974}

\bibitem[{Unger {et~al.}(2023)Unger, Ségransan, Barbato, Delisle, Sahlmann, Holl, \& Udry}]{unger_exploring_2023}
Unger, N., Ségransan, D., Barbato, D., {et~al.} 2023, Astronomy \& Astrophysics, 680, A16, \dodoi{10.1051/0004-6361/202347578}

\bibitem[{Ward-Duong {et~al.}(2015)Ward-Duong, Patience, De~Rosa, Bulger, Rajan, Goodwin, Parker, McCarthy, \& Kulesa}]{ward-duong_m-dwarfs_2015}
Ward-Duong, K., Patience, J., De~Rosa, R.~J., {et~al.} 2015, Monthly Notices of the Royal Astronomical Society, 449, 2618, \dodoi{10.1093/mnras/stv384}

\bibitem[{Xiao {et~al.}(2023)Xiao, Liu, Teng, Wang, Brandt, Zhao, Zhao, Zhai, \& Gao}]{xiao_masses_2023}
Xiao, G.-Y., Liu, Y.-J., Teng, H.-Y., {et~al.} 2023, Research in Astronomy and Astrophysics, 23, 055022, \dodoi{10.1088/1674-4527/accb7e}

\bibitem[{Yang {et~al.}(2020)Yang, Zhang, Hwang, Zang, Gould, Wang, Mao, Albrow, Chung, Han, Jung, Ryu, Shin, Shvartzvald, Yee, Zhu, Penny, Fouqué, Cha, Kim, Kim, Kim, Lee, Lee, Lee, Park, \& Pogge}]{yang_kmt-2016-blg-1836lb_2020}
Yang, H., Zhang, X., Hwang, K.-H., {et~al.} 2020, The Astronomical Journal, 159, 98, \dodoi{10.3847/1538-3881/ab660e}

\bibitem[{Zandt {et~al.}(2025)Zandt, Gilbert, Giacalone, Petigura, Howard, \& Handley}]{zandt_smooth_2025}
Zandt, J.~V., Gilbert, G., Giacalone, S., {et~al.} 2025, A {Smooth} {Transition} from {Giant} {Planets} to {Brown} {Dwarfs} from the {Radial} {Occurrence} {Distribution},  arXiv, \dodoi{10.48550/arXiv.2511.18758}

\bibitem[{Zang {et~al.}(2021)Zang, Hwang, Udalski, Wang, Zhu, Sumi, Yee, Gould, Mao, Zhang, Authors), Albrow, Chung, Han, Jung, Ryu, Shin, Shvartzvald, Cha, Kim, Kim, Kim, Lee, Lee, Lee, Park, Pogge, Collaboration), Mróz, Skowron, Poleski, Szymański, Soszyński, Pietrukowicz, Kozłowski, Ulaczyk, Rybicki, Iwanek, Wrona, Gromadzki, Collaboration), Bond, Abe, Barry, Bennett, Bhattacharya, Donachie, Fujii, Fukui, Hirao, Itow, Kirikawa, Kondo, Koshimoto, Li, Matsubara, Muraki, Miyazaki, Olmschenk, Ranc, Rattenbury, Satoh, Shoji, Silva, Suzuki, Tanaka, Tristram, Yamawaki, Yonehara, Collaboration), Beichman, Bryden, Novati, Carey, Gaudi, Henderson, Johnson, \& Team)}]{zang_systematic_2021}
Zang, W., Hwang, K.-H., Udalski, A., {et~al.} 2021, The Astronomical Journal, 162, 163, \dodoi{10.3847/1538-3881/ac12d4}

\bibitem[{Zang {et~al.}(2022)Zang, Yang, Han, Lee, Udalski, Gould, Mao, Zhang, Zhu, Albrow, Chung, Hwang, Jung, Ryu, Shin, Shvartzvald, Yee, Cha, Kim, Kim, Kim, Lee, Lee, Park, Pogge, Mróz, Skowron, Poleski, Szymański, Soszyński, Pietrukowicz, Kozłowski, Ulaczyk, Rybicki, Iwanek, Wrona, \& Gromadzki}]{zang_systematic_2022}
Zang, W., Yang, H., Han, C., {et~al.} 2022, Monthly Notices of the Royal Astronomical Society, 515, 928, \dodoi{10.1093/mnras/stac1883}

\bibitem[{Zang {et~al.}(2023)Zang, Jung, Yang, Zhang, Udalski, Yee, Gould, Mao, Authors), Albrow, Chung, Han, Hwang, Ryu, Shin, Shvartzvald, Cha, Kim, Kim, Kim, Lee, Lee, Lee, Park, Pogge, Collaboration), Mróz, Skowron, Poleski, Szymański, Soszyński, Pietrukowicz, Kozłowski, Ulaczyk, Rybicki, Iwanek, Wrona, Gromadzki, Collaboration), Wang, Zhang, Zhu, \& Collaboration)}]{zang_systematic_2023}
Zang, W., Jung, Y.~K., Yang, H., {et~al.} 2023, The Astronomical Journal, 165, 103, \dodoi{10.3847/1538-3881/acb34b}

\bibitem[{Zang {et~al.}(2025)Zang, Jung, Yee, Hwang, Yang, Udalski, Sumi, Gould, Mao, Albrow, Chung, Han, Ryu, Shin, Shvartzvald, Cha, Kim, Kim, Kim, Lee, Lee, Lee, Park, Pogge, Zhang, Kuang, Wang, Zhang, Hu, Zhu, Mróz, Skowron, Poleski, Szymański, Soszyński, Pietrukowicz, Kozłowski, Ulaczyk, Rybicki, Iwanek, Wrona, Gromadzki, Abe, Barry, Bennett, Bhattacharya, Bond, Fujii, Fukui, Hamada, Hirao, Silva, Itow, Kirikawa, Koshimoto, Matsubara, Miyazaki, Muraki, Olmschenk, Ranc, Rattenbury, Satoh, Suzuki, Tomoyoshi, Tristram, Vandorou, Yama, \& Yamashita}]{zang_microlensing_2025}
Zang, W., Jung, Y.~K., Yee, J.~C., {et~al.} 2025, Science, 388, 400, \dodoi{10.1126/science.adn6088}

\bibitem[{Zhang {et~al.}(2024)Zhang, Zang, El-Badry, Lu, Bloom, Agol, Gaudi, Konopacky, LeBaron, Mao, \& Terry}]{zhang_earth-mass_2024}
Zhang, K., Zang, W., El-Badry, K., {et~al.} 2024, Nature Astronomy, 8, 1575, \dodoi{10.1038/s41550-024-02375-9}

\end{thebibliography}

\bibliographystyle{aasjournal}
\end{CJK*}
\end{document}